\documentclass[prc,aps,showpacs]{revtex4}
\usepackage{graphicx}
\begin{document}
%\begin{frontmatter}
\title
{An attempt to explore the production routes of Astatine radionuclides: Theoretical approach}

\author{Moumita Maiti}
\affiliation{Saha Institute of Nuclear Physics, 1/AF, Bidhannagar, Kolkata-700064, India.}
\author{ Susanta Lahiri}
\affiliation{Saha Institute of Nuclear Physics, 1/AF, Bidhannagar, Kolkata-700064, India.}

\begin{abstract}
In order to fulfill the recent thrust of Astatine radionuclides in the field of nuclear medicine various production routes have been explored in the present work. The possible production routes of $^{209-211}$At comprise both light and heavy ion induced reactions at the bombarding energy range starting from threshold to maximum 100 MeV energy. For this purpose, we have used the nuclear reaction model codes TALYS, ALICE91 and PACE-II. Excitation functions of those radionuclides, produced through various production routes, have been calculated using nuclear reaction model codes and compared with the available measured data. Contribution of various reaction mechanisms, like, direct, preequilibrium and equilibrium reactions, to the total reaction cross section has been studied using the codes. Result shows that equilibrium reaction mechanism dominates in all cases over other reaction mechanisms.

%\begin{keyword}
%excitation function, heavyion induced reactions, $\alpha$-induced reaction, nuclear reaction model codes
\end{abstract}
%\end{keyword}
%\end{frontmatter}
\maketitle

\section{Introduction}

In the last decade semi-metallic radioactive element astatine (At) has drawn ample attention in the field of nuclear medicine. Nuclear physics has an important role in developing this field as nuclear reaction is the method of choice to produce At radionuclides. Most of the identified radioisotopes of At are short lived. $^{210}$At, which has half life, T$_{1/2}$=8.1 h, is the most stable isotope among others. Though At radioisotopes are produced artificially, however, in the Earth's crust, small amount of At may last in equilibrium with uranium and thorium. At was discovered about 80 years ago, but, studies on its chemical properties faced practical constraint due to its scarcity and short half life. Some recent reports on chamemical separation of At radionuclides from the bulk, which is also an important step followed before application, has been cited in ref \cite {1a,1b,1c}. A list of At radionuclides, which may find possible use in nuclear medicine, having suitable half life is presented in Table \ref{mmt1}.

\begin{table}
\caption{The nuclear properties \cite{Tiso} of At radionuclides useful in the field of nuclear medicine}
\label{mmt1}
\begin{tabular}{cccccc}
\hline
     Radionuclide & T$_{1/2}$ & Decay mode & $\alpha$(MeV) & $\gamma$(keV)(\%)\\
\hline
$^{211}$At & 7.214 $h$ & $\epsilon$ 58.2\%, $\alpha$ 41.8\% &  5.867 & 687(0.246)\\
$^{210}$At  & 8.1 $h$ & $\epsilon$ 99.82\%, $\alpha$ 0.18\% &  5.524,5.361 & 1181.43(99.3),245.35(79),1483.35(46.5)  \\
$^{209}$At & 5.41 $h$ & $\epsilon$ 95.9\%, $\alpha$ 4.1\%  &  5.647 & 545.03(91),781.89(83.5), 790.2(63.5)  \\
$^{208}$At & 1.63 $h$ & $\epsilon$ 99.45\%, $\alpha$ 0.55\% &  5.641 & 685.2(97.89),660.1(90), 177.0(46)  \\

\hline
\end{tabular}
\end{table}

Among others, $^{211}$At is most promising radionuclide for targeted therapy due to its suitable half life and relatively high linear energy transfer (LET) of $\alpha$-particles in human tissues. $^{211}$At has two major decay modes both of which eventually emits high energy $\alpha$-particles suitable for targetted therapy. The $\alpha$-decay mode emits $\alpha$-particle of energy 5.867 MeV leaving $^{207}$Bi (T$_{1/2}$=32.9 y) as residual, which decays to stable $^{207}$Pb. Other decay mode produces ultra short lived daughter, $^{211g}$Po (T$_{1/2}$ = 516 ms, J =9/2$^{+}$), which decays to stable $^{207}$Pb emitting 7.45 MeV $\alpha$-particle along with the emission of X-rays. The metastable state of $^{211m}$Po (T$_{1/2}$ = 25.2 s, J =25/2$^{+}$) is not populated by $^{211}$At because of its high spin and low transition probabilities. The  $\gamma$-emitting radionuclides $^{209,210}$At have mostly $\epsilon$-capture decay mode and produce long lived $\alpha$-emitting $^{209,210}$Po daughter. These At radionuclides may find  uses in diagnosys due to their suitable half life and high intensity $\gamma$-rays.

Radionuclidic purity is an important issue in nuclear medicine. Therefore, At radionuclide for medical application has to be obtained in its purest form to avoid unnecessary radiation exposure from impurities. In general, production route and the chemical separation technique are the two major sources of impurity (radionuclides as well as stable) that is propagated to the particular isotope of interest during its production and purification procedures. Production route hosts radionuclidic impurity and their corresponding decay products including the decay of compound nucleus. Optimization of nuclear reaction parameters, like, suitable target-prijectile combination, bombarding energy of projectile, thickness of target, etc., is a probable way to reduce certain production of impurities. For example, the most commonly used production route of At radionuclides is $^{209}$Bi($\alpha$,xn)$^{209-211}$At \cite{1,2,3,4,5,6,7}. These At isotopes and their daughter products (radioactive and stable) can be considered as impurity when aimed for a particular isotope. Long lived impurities are hazardous to human health if the amount goes above a certain level while the stable heavy metals, Pb, Tl, which are the end products of At radioisotopes, are sometimes carcinogenic and affects human metabolism. 

At radionuclides can be produced directly by nuclear reactions using suitable target-projectile combination or from the decay of its precursor Rn isotopes. Light and heavy ion induced reactions may be used for the production of At radionuclides in low and medium energy accelerators. High energy proton induced spallation reactions on heavy targets like, $^{238}$U, $^{232}$Th can also be used to produce $^{209-211}$At. In the indirect method, $^{209-211}$Rn (T$_{1/2}$=28.5 m, 2.4 h and 14.6 h respectively) isotopes are produced either by light charged particle induced reactions or by spallation reaction induced by high energy proton. At isotopes are produced as a daughter product of Rn isotopes. Various production routes of $^{209-211}$At escort different type of impurities whose quantification is important for their application in nuclear medicine.

The present work is to explore the possible production routes of $^{209-211}$At radionuclides from both light and heavy ion induced reactions. Production of At radionuclides is preferred from $\alpha$-induced reaction on $^{209}$Bi because of high yield required for theraputic purpose. However, productions via light heavy ion induced reactions may serve some other scientific purposes,like, biological studies which requires minute amount of radioactivity. We aim in this paper to investigate theoretically the various possible production routes of $^{209-211}$At from both light and heavy ion induced reactions and the reaction mechanisms involved in each route. A comparison of theoretically predicted production cross sections has been made with the reported experimental values wherever available. 

Production of $^{209-211}$At radionuclides from light ($\alpha$ and $^{3}$He) and heavy ion ($^{6,7}$Li, $^{9}$Be) induced reactions on heavy natural targets like Bi, Pb, Tl, involves mostly preequilibrium (PEQ) and equilibrium (EQ) reaction mechanisms in the low and medium energy range. In case of high projectile energy, direct reaction (DIR) mechanism may also have some contribution. A composite nucleus is formed when a target is bombarded by the projectile of certain energy. Particle emission may occur from any stage of the relaxation process of the target-projectile composite system leaving the residual in an excited state. The excited residual nucleus is cooled either by particle emission, if the residual excitation is higher than the particle emission threshold giving birth to a new residual nucleus, or by gamma ray emission. 

In case of direct reaction, particles are emitted due to large momentum transfer from the very first interaction between the projectile and the target. As a result ejectile carries high energy leaving the residual at lower excitation. When incoming energy of the projectile is shared by all the nucleons of the composite system, equilibrium condition is reached. Particles are emitted from the compound nucleus due to statistical fluctuation in energy. Therefore, compound nuclear emissions are low energy emissions and the residual is left at high excitation energy. 

Between the two extremes, there is an intermediate stage of reaction which exhibits both direct and compound like features. These reactions are referred to as PEQ processes. PEQ emissions take place after the first stage of the reaction (DIR) but long before statistical equilibrium of the compound nucleus is attained. PEQ processes cover a significant part of the reaction cross section for incident energies between 10 and 200 MeV.
The production cross section of the desired radionuclide is the resultant of all the reaction mechanisms involved. However, contribution of direct reaction mechanism is comparatively small in the energy range up to 100 MeV considered for the production of At-radionuclides in the present work.
%We have calculated excitation functions of $^{209-211}$At radionuclides produced via $^{209}$Bi($\alpha$,xn), $^{nat}$Pb($^{7}$Li, xn), $^{nat}$Pb($^{6}$Li, xn), $^{nat}$Tl($^{9}$Be,xn) reactions using the nuclear reaction model codes TALYS, ALICE91, PACE-II wherever suitable.

Light and heavy ion induced nuclear reactions have been studied using the nuclear reaction model codes TALYS \cite{8}, ALICE91 \cite{9, 10} and PACE-II \cite{11}. The code TALYS takes into account all three types of reaction mechanisms; direct (Coupled channel analysis, Distorted Wave Born Approximation (DWBA) etc.), preequilibrium (one and two component exciton model) and equilibrium (Hauser-Feshbach model), PACE-II accounts only evaporation reactions based on Hauser-Feshbach model \cite{12} and ALICE91considers preequilibrium(hybrid model) along with the equilibrium reactions (Weisskopf-Ewing model) \cite{13}. A list of reactions studied in the present work is presented in Table \ref{mmt2}. Use of three codes enables us to understand the reliability as well as predictablity of the codes.

In the next section we give a brief description of the codes TALYS, ALICE91 and PACE-II and the input option used for the present calculation. Sec III shows the remarkable features of the codes. In sec IV we discuss the results obtained in the present study.
\begin{table}
\caption{List of nuclear reactions studied along with corresponding $Q$ values and threshold energies}
\label{mmt2}
\begin{tabular}{ccc}
\hline
   Reaction & $Q$ value (MeV) & E$_{th}$(MeV) \\
\hline
 $^{209}$Bi($\alpha$, $2n$)$^{211}$At & -20.33 & 20.72  \\
 $^{209}$Bi($\alpha$, $3n$)$^{210}$At & -28.07 & 28.61 \\
 $^{209}$Bi($\alpha$, $4n$)$^{209}$At & -35.24 & 35.92  \\
 $^{209}$Bi($^{3}$He, $n$)$^{211}$At & 0.25 & 0  \\
 $^{209}$Bi($^{3}$He, $2n$)$^{210}$At & -7.50 & 7.61 \\
 $^{209}$Bi($^{3}$He, $3n$)$^{210}$At & -14.66 & 14.87 \\
 $^{208}$Pb($^{6}$Li, $3n$)$^{211}$At & -20.23 & 20.81 \\
 $^{208}$Pb($^{6}$Li, $4n$)$^{210}$At & -27.98 & 28.78 \\
 $^{208}$Pb($^{6}$Li, $5n$)$^{209}$At & -35.14 & 36.15 \\
 $^{208}$Pb($^{7}$Li, $4n$)$^{211}$At & -27.48 & 28.41 \\
 $^{208}$Pb($^{7}$Li, $5n$)$^{210}$At & -35.23 & 36.41 \\
 $^{208}$Pb($^{7}$Li, $6n$)$^{209}$At & -42.39 & 43.82 \\
 $^{203}$Tl($^{9}$Be, $n$)$^{211}$At & -10.84 & 11.32 \\
 $^{203}$Tl($^{9}$Be, $2n$)$^{210}$At & -18.58 & 19.41 \\
 $^{203}$Tl($^{9}$Be, $3n$)$^{210}$At & -25.75 & 26.89 \\
 $^{205}$Tl($^{9}$Be, $3n$)$^{211}$At & -25.04 & 26.14 \\
 $^{205}$Tl($^{9}$Be, $4n$)$^{210}$At & -32.79 & 34.23 \\
 $^{205}$Tl($^{9}$Be, $5n$)$^{209}$At & -39.95 & 41.71 \\

\hline
\end{tabular}
\end{table}

\section{Brief description of nuclear reaction model codes}

\subsection{ALICE91}

We have used the code ALICE91 \cite{9, 10} to calculate the excitation function of At radionuclides produced through various reaction mechanisms. Geometry dependent hybrid model \cite{10, 14,15,16} has been used to calculate PEQ emissions and Weisskopf-Ewing formalism \cite{13} for EQ emissions. ALICE91 does not take care of DIR emissions. 
The hybrid model is the combination of exciton model \cite{17} and Boltzmann master equation approach \cite{18, 19}. It assumes that the target-projectile composite system proceeds through two body interaction process. Each stage of the relaxation process is designated by the total number (n) of excited particles, sum of excited particles (p) and hole (h). In each two body interaction, p-h pair may be created or annihilated or redistribution of energy takes place without changing the number. Hybrid model uses "never come back" approximation, i.e, the model assumes only p-h pair is created in each interaction. Hybrid model explicitly determines the pre-emission energy distribution of the excited particles which helps to estimate high energy emissions more accurately. Geometry dependent hybrid model includes the nuclear surface effects \cite{10, 15, 16}. The PEQ emission cross section for a particular ejectile $x$ with energy $\epsilon_{x}$ is given by

\begin{equation}
\label{mmeq1}
\begin{array}{lll}
\sigma _{PEQ}(\epsilon _{x}) & = 
 & \frac{\lambda ^{2}}{4\pi}\sum\limits_{l=0}^{\infty}(2l+1)T_{l}\sum  \limits
^{\overline{n}}_{n=n_{0},\atop {\Delta n=2}}D_{n}\left[f^{x}_{n}\frac{N_{n}(l,U,\epsilon_{x})}{N_{n}(l,E_{c})}\right]{\lambda _{c}(\epsilon _{x})\over{\lambda _{c}(\epsilon _{x})+\lambda _{t}(\epsilon _{x})}}
\end{array}
\end{equation}

Here, $\lambda$ is the de-Broglie wave length of the projectile, $T_{l}$ is the transmission coefficient of the $l^{th}$ partial wave, $D_{n}$ is the depletion factor of the $n^{th}$ exciton state, that is, probability of reaching $n$ exciton state without prior emission and $f^{x}_{n}$   is number of $x$ type excited nucleon present in it. The number $n_{0}$ and $\overline n$  are the initial and equilibrium exciton numbers respectively. The ratio ${N_{n}(l,U,\epsilon_{x})}/{N_{n}(l,E_{c})}$  is the probability of finding $x$ type nucleon in the $n$ exciton state with energy ($\epsilon_{x}+B_{x}$)  where $B_{x}$ is the separation energy of $x$.  The factor  ${\lambda _{c}(\epsilon 
_{x})\over{\lambda _{c}(\epsilon _{x})+\lambda _{t}(\epsilon _{x})}}$  is the emission probability of $x$ with energy $\epsilon_{x}$. $\lambda _{t}(\epsilon _{x})$  is the two-body interaction rate. The emission rate $\lambda _{c}(\epsilon _{x})$  is calculated by \cite{10}
\begin{equation}
\label{mmeq2}
\lambda_{c}(\epsilon _{x})=\frac{(2S_{x}+1)\mu_{x}\epsilon_{x}\sigma_{inv}(\epsilon_{x})}{\pi^{2}\hbar^{3}g} 
\end{equation} 
where, $S_{x}$  is the intrinsic spin of $x$, $\mu_{x}$   is the reduced mass, $\sigma_{inv}$  is the inverse cross section of the ejectile $x$ with energy $\epsilon_{x}$  being absorbed by the residual and $g$  is the single particle level density of the composite nucleus. 
Equilibrium emission cross section is calculated using Weisskopf-Ewing formalism as
 \begin{equation}
\label{mmeq3}
\sigma_{EQ}(\epsilon _{x}) \sim \sigma_{comp}\frac{e^{2(aU)^{1/2}}}{U}
\end{equation}
$\sigma_{comp}$  is the compound nuclear formation cross section, $a$ is the level density parameter and $U$ is the available excitation energy of the compound nucleus after the preequilibrium emissions. $\sigma_{comp}$  is calculated as  $\sigma_{comp}=\sigma_{abs}-\sigma_{PEQ}$ where $\sigma_{abs}$  is the absorption cross section of the projectile in the target and $\sigma_{PEQ}$  is the total preequilibrium emission cross section. 
  
\subsubsection{Input options used}

The calculation has been performed using the code ALICE91 \cite{9, 10} with geometry dependent hybrid model for PEQ emissions and Weisskopf-Ewing formalism for EQ emissions. Neutron, proton, alpha and deuteron emissions are considered from the residual nuclides of 12 mass unit wide and 10 charge unit deep including the composite nucleus. Fermi gas level density has been used for the calculation of reaction cross section. Reverse channel reaction cross sections have been calculated using the optical model. Default value of the level density parameter, $A/9$ has been chosen for the calculation. Total number of nucleons in the projectile has been chosen as the initial exciton number for the cross section calculation.

\subsection{TALYS}
The code TALYS \cite{8} has been used to calculate the production of At radionuclides through $\alpha$-particle and $^{3}$He-induced reactions on $^{209}$Bi target. We have adopted two-component exciton model for the estimation of PEQ emissions, detail Hauser-Feshbach formalism \cite{12} for the EQ emissions and Coupled Channel analysis for DIR reaction process. In case of PEQ reactions of projectiles and ejectiles heavier than nucleon ($d$, $t$, $^{3}$He and $\alpha$-particles) direct like reaction processes; stripping, pickup and knockout reactions have important contribution in addition to the exciton model and are treated in the frame work of Kalbach's phenomenological model \cite{19a} as these processes are not included in the exciton model.
 
In the two-component exciton model, total exciton number is divided into proton ($p_{p}$) /neutron ($p_{n}$)  particle number and proton ($h_{p}$) /neutron ($h_{n}$)  hole number. $n_{p}(=p_{p}+h_{p})$ and $n_{n}(=p_{n}+h_{n})$  are the proton and neutron exciton numbers respectively. PEQ cross section of the particle $x$ with energy $\epsilon_{x}$  is calculated as
\begin{equation}
\label{mmeq4}
\begin{array}{lll}
\sigma_{PEQ}(\epsilon _{x})& = & \sigma^{CF}\sum\limits^{\overline p_{p}}_{p_{p}=p^{0}_{p}}\sum\limits^{\overline p_{n}}_{p_{n}=p^{0}_{n}}D(p_{p},h_{p},p_{n},h_{n})W(p_{p},h_{p},p_{n},h_{n},\epsilon_{x})\tau(p_{p},h_{p},p_{n},h_{n})
\end{array}
\end{equation}

where, $\sigma^{CF}$   is the composite nucleus formation cross section, $D(p_{p},h_{p},p_{n},h_{n})$  is a fraction of PEQ population that has survived previous  emission, $W(p_{p},h_{p},p_{n},h_{n},\epsilon_{x})$  is the emission rate of $x$ with energy $\epsilon_{x}$   and $\tau$  is the mean life time of that state. 
Compound nuclear emission is calculated in TALYS by two different mechanisms; binary compound cross section that is the capture of the projectile in the target nucleus subsequently emitting particle or $\gamma$-ray at low energy and multiple compound emissions, that is, multiple emissions of highly excited residual nuclei formed after the binary reaction. Width fluctuation is taken into account in case of binary compound cross section. Multiple compound and precompound emissions are considered in the framework of Harser-Feshbach and exciton model. TALYS contains both macroscopic and microscopic level densities. Macroscopic level densities are\\ 
(i) Constant Temperature model \cite{20} and Fermi gas model \\
(ii) Back-shifted Fermi gas model and \\
(iii) generalized superfluid model.

\subsubsection{Input options used}

The code TALYS uses detail Hauser-Feshbach formalism for multiple particle emission in the evaporation calculation along with width fluctuation corrections using Moldauer model \cite{21, 22}. The option for multiple pre-equilibrium emissions has been chosen within the frame work of two-component exciton model calculation with surface correction to account the pre-equilibrium emissions. We have calculated level densities using a combination of the Constant Temperature Model by Gilbert and Cameron \cite{20} and the Fermi gas model. The total excitation energy range is divided into two regions: low energy part and high energy part. Low energy part starts from 0 MeV to the certain energy, $E_{mid}$, up to which constant temperature law is valid. The high energy part starts above the $E_{mid}$ and Fermi gas model is used to calculate level densities in this zone.

\subsection{PACE-II}

The code PACE-II is the modified version of the Monte-Carlo code Projection Angular-momentum Coupling Evaporation. The deexcitation process of the excited nuclei is calculated using the modified version of the code JULIAN \cite{23}, which follows the correct procedure of angular momentum coupling at each stage of deexcitation.
The code can be used to study the high energy heavy ion reactions. The transmission coefficients for light particle emission are determined from the optical model potential where all the optical model parameters are kept as default. The shift in coulomb barrier during deexcitation is accounted by calculating the transmission coefficients at an effective energy determined by the shift. The code internally decides level densities and masses it needs during deexcitation. Gilber-Cameron level density is taken as default. Fission is considered as a decay mode where the fission barrier can be changed accordingly in the program. The default fission barrier is the Cohen-Plasil-Swiatecki rotating liquid drop fission barrier. Saddle point level density is proportional to the $a$ parameter. For compound nucleus reactions distributions are determined as function of angle around recoil axis and for deep inelastic products, angular distributions are determined as functions of the angle axis perpendicular to the reaction plane. Compound nuclear fusion cross section is determined by using the Bass method \cite{24}. A non-statistical yrast cascade $\gamma$-decay chain has been artificially incorporated to simulate $\gamma$ multiplicity and energy results. 

\section{	Important features of the codes}

(1)	TALYS uses exciton model to calculate PEQ emissions followed by Hauser-Feshbach formalism for EQ emissions. ALICE91 uses hybrid model for PEQ and Weisskopf-Ewing formalism for EQ emissions\\
(2)	TALYS relies on DIR emissions treated by coupled channel analysis,  DWBA \cite{25}, Giant resonance \cite{26, 27}, etc.  However, ALICE91 does not consider DIR emissions.\\
(3)	Direct like processes, for example, stripping, pickup, knockout, etc., have important role in the nuclear reactions involving cluster ($d$, $t$, $^{3}$He and $\alpha$-particles) as projectile and ejectile. TALYS includes these processes in the frame work of Kalbach's phenomenological model. ALICE91 does not take into account the above described processes.\\
(4)	TALYS considers $n$, $p$, $d$, $t$, $^{3}$He and $\alpha$-particle as both projectile and ejectile with single and simultaneous emission of any order in PEQ and EQ emissions. ALICE91 considers light particles up to $\alpha$ as projectile but only $n$, $p$, $d$ and $\alpha$-particle are considered as ejectiles. Single and simultaneous emission of two nucleons are taken into account in PEQ emissions and in case of EQ reactions, $n$, $p$, $d$ and $\alpha$-particle emissions are calculated from successive residuals. PEQ emissions from heavy ion induced reactions can also be calculated using hybrid model in ALICE91with necessary modifications.\\
(5)	PACE-II is a Monte-Carlo statistical code based on Hauser-Feshbach formalism and the decay channels consider statistical emission of $\gamma$-rays, $n$, $p$ and $\alpha$-particles.

\section{	Results and Discussions}

In the present report excitation function of $^{209-211}$At have been calculated using the reaction model codes TALYS, ALICE91 and PACE-II produced through various possible target projectile combination in the incident energy range starting from threshold to 100 MeV. Table \ref{mmt2} represents the details of reactions with the reaction threshold values. Production of $^{209-211}$At through $^{209}$Bi($\alpha$, $xn$) and $^{209}$Bi($^{3}$He, $xn$) reactions have been studied using the code TALYS and ALICE91 and have compared with the available experimental cross section. Other possible production roots of At radionuclides via heavy ion induced reactions have been studied using the codes ALICE91 and PACE-II and have also compared with the available measured data. It has been seen that all the measured excitation functions of $^{211}$At \cite{1, 31,32,33} produced through $^{209}$Bi($\alpha$, 2$n$)$^{211}$At reaction are in good agreement with each other.

It has been seen earlier that $\alpha$-particle may interact with the target nucleons in different ways. Among others, 95\% contribution comes from the two modes of interaction to the total reaction cross section; (i) $\alpha$-particle dissolves into the four constituent nucleons in the nuclear force field ($n_{0}$=4; 4$p$-0$h$) \cite{28, 29} (ii) $\alpha$-particle as a whole interacts with a target nucleon and creates a particle-hole pair that is the starting point of the PEQ cascade ($n_{0}$=6; 5$p$-1$h$). It has also been found that both these initial configurations have proportional contribution to the angular distribution of the ejectiles \cite{mm}.
ALICE91 has option to select initial exciton number, $n_{0}$ in the input. We have performed separate calculation of excitation function for both the options $n_{0}$=4 and 6.  The resultant excitation functions of $^{209-211}$At have been calculated taking the 70\% contribution from $n_{0}$=6 and 30\% of that from $n_{0}$=4 according to the observation of Gadioli {\it et al} \cite{30}. 

\begin{figure}
\begin{center}
\includegraphics[height=8.0cm]{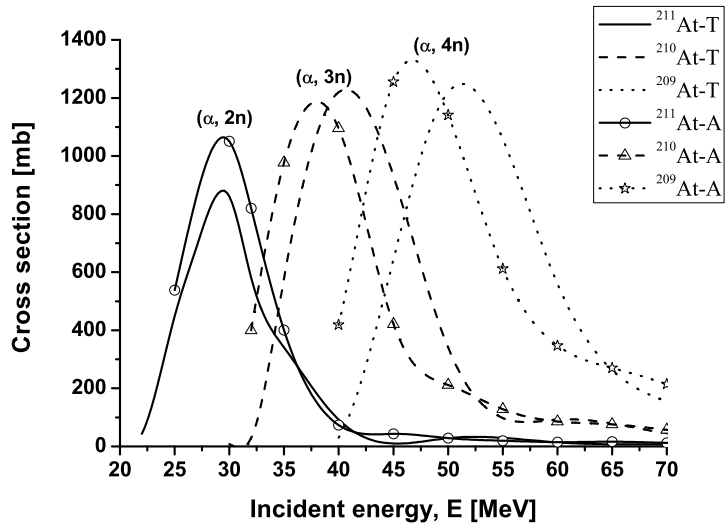}
\caption{Comparison between the calculated excitation functions of $^{209-211}$At-radionuclides from the reaction $^{209}$Bi($\alpha$, $xn$)$^{209-211}$At using the codes TALYS (lines) and ALICE91(lines with symbols).} 
\label{fig1}
\end{center}
\end{figure}

\begin{figure}
\begin{center}
\includegraphics[height=8.cm]{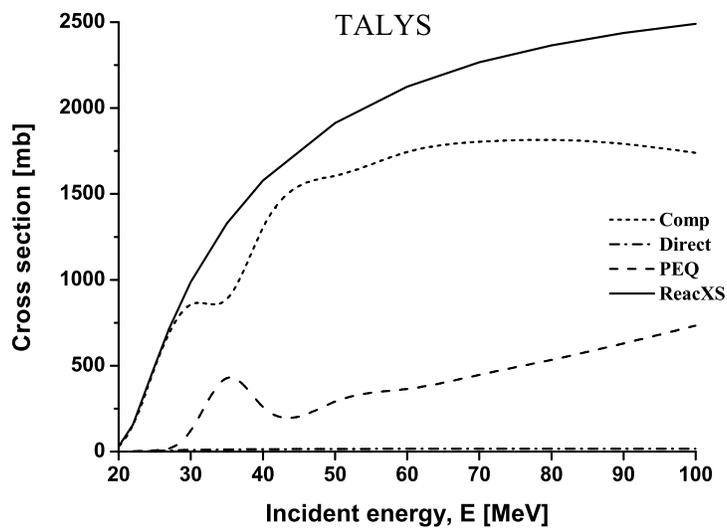}
\caption{Contributions of different reaction mechanisms to the total reaction cross section as obtained from TALYS.} 
\label{fig2}
\end{center}
\end{figure}

\begin{figure}
\begin{center}
\includegraphics[height=8.cm]{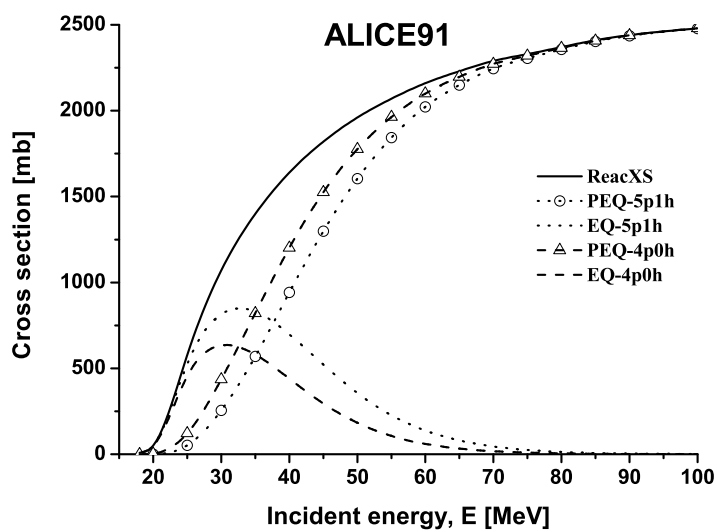}
\caption{Contribution of total PEQ and EQ reactions to the total cross section as obtained from ALICE91 for two initial exciton numbers ($n_{0}$=4; 4$p$-0$h$ and $n_{0}$=6; 5$p$-1$h$) in the $\alpha$-induced reaction.} 
\label{fig3}
\end{center}
\end{figure}

Figure \ref{fig1} shows a comparison between the formation cross section of $^{209-211}$At produced from $^{209}$Bi via ($\alpha,2n$), ($\alpha,3n$) and ($\alpha,4n$) reaction channels calculated using TALYS and ALICE91. In the TALYS calculation, optical model has been used to calculate inverse reaction cross section with default level density option, which is the combination of Constant Temperature model of Gilbert and Cameron \cite{20} and Fermi gas model. In the ALICE91 calculation we have used optical model with Fermi gas level density option. All three isotopes of At are well separated from each other and the maximum production of $^{211}$At and $^{210}$At have also been seen at the at the incident energies, 30 MeV and 40 MeV respectively in both the results. But, ALICE91 produces 20\% more $^{211}$At and 10\% less $^{210}$At than TALYS. Maximum production of $^{209}$At has been seen at 45 MeV in ALICE91 and that at 50 MeV in TALYS result. 

To investigate the results obtained from the two model codes, we have studied the contributions from the reaction mechanisms involved in $\alpha$-induced reaction on $^{209}$Bi target. Contributions from PEQ, EQ and DIR reactions to the total reaction cross section have been shown against incident projectile energy from TALYS and ALICE separately in Figure \ref{fig2} and \ref{fig3}. In figure \ref{fig2}, we see that there is almost no contribution from direct reaction for incident energy up to 100 MeV. No PEQ reaction contribution has been seen in low incident energies, however, PEQ component increases with the increasing incident energy. The most dominating component is the EQ reaction in the range considered for the present work. It follows similar increasing trend up to 50 MeV with the total reaction cross section, which shows increasing trend through the entire energy range considered. The rate of increment of EQ cross section is slow at the higher incident energies as PEQ component is increasing with the increasing incident energy. In case of ALICE91 in figure \ref{fig3}, we see that EQ contribution is significant only at low energies for both the initial configurations (4$p$-0$h$ \& 5$p$-1$h$) of $\alpha$-induced reaction. PEQ component is increasing rapidly with the increasing incident energy for both the initial exciton numbers. The total cross section also shows the increasing trend throughout the range. PEQ contribution from 5$p$-1$h$ configuration is less than that of 4$p$-0$h$ configuration of $\alpha$ up to the energy range 60 MeV.  This is due to the fact that in case of 5$p$-1$h$ configuration, PEQ emission starts in the hybrid model calculation from $n_{0}$=6, which is one stage later that $n_{0}$=4. As a result, PEQ contribution is larger for $n_{0}$=4 than the other. From the plot we see that total PEQ cross section does not increase above 60 MeV for both the cases. This is because, in the code ALICE91  has a cut off value (= $n_{0}$+18) if it exceeds this value.  The case of EQ reaction is just reverse between $n_{0}$=4 and $n_{0}$=6. This is because after sufficient preequilibrium emission from the 4$p$-0$h$ configuration total available excitation becomes low which in turn reduces the EQ emission.

\begin{figure}
\begin{center}
\includegraphics[height=8.cm]{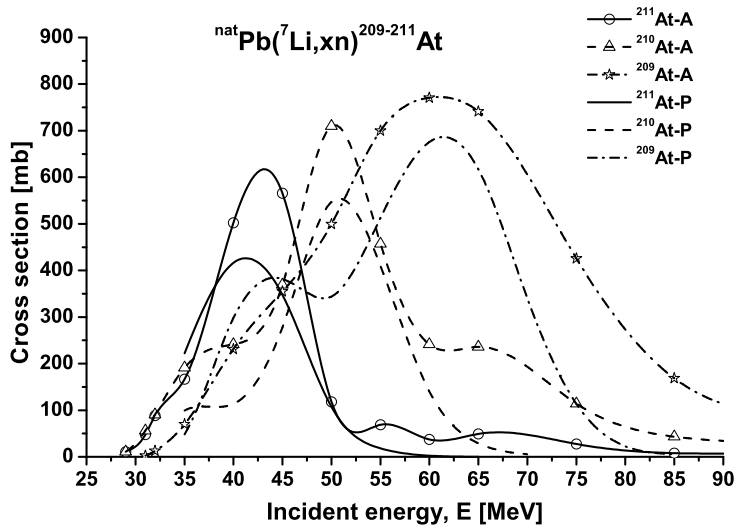}
\caption{Comparison between the formation cross sections of $^{209-211}$At from $^{7}$Li induced reaction on $^{nat}$Pb target using ALICE91 and PACE-II.} 
\label{fig4}
\end{center}
\end{figure}

\begin{figure}
\begin{center}
\includegraphics[height=8.cm]{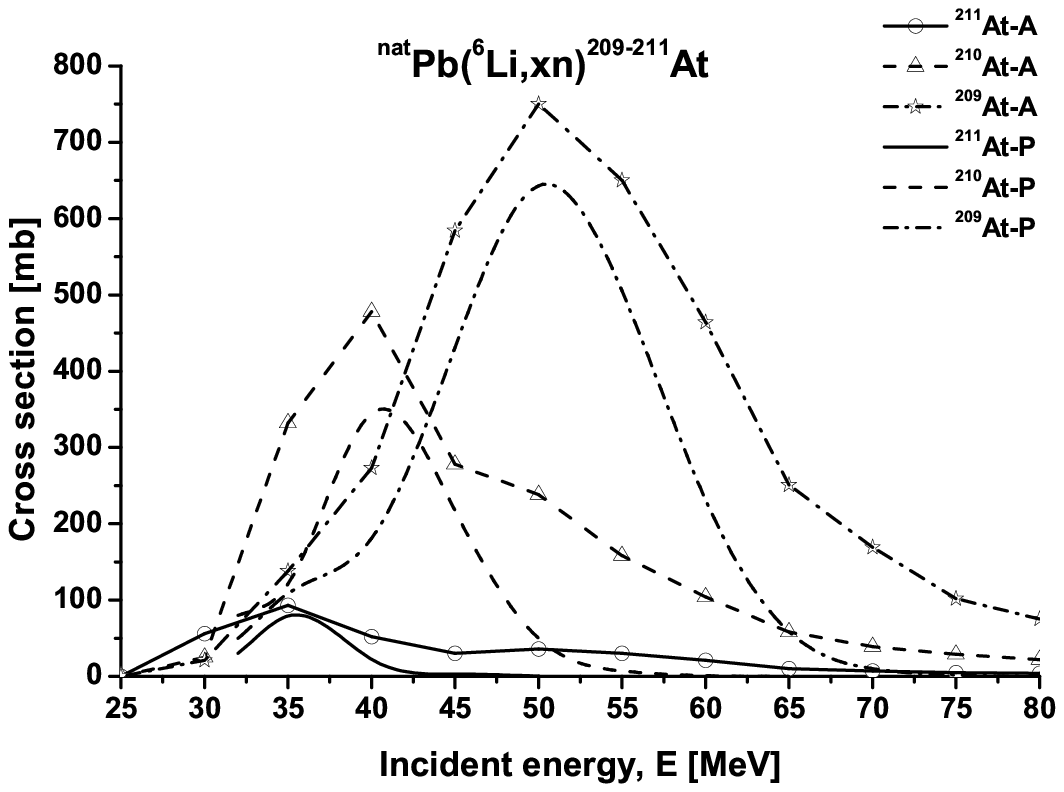}
\caption{Comparison between the formation cross sections of $^{209-211}$At from $^{6}$Li-induced reaction on $^{nat}$Pb target using ALICE91 and PACE-II.} 
\label{fig5}
\end{center}
\end{figure}

\begin{figure}
\begin{center}
\includegraphics[height=8.cm]{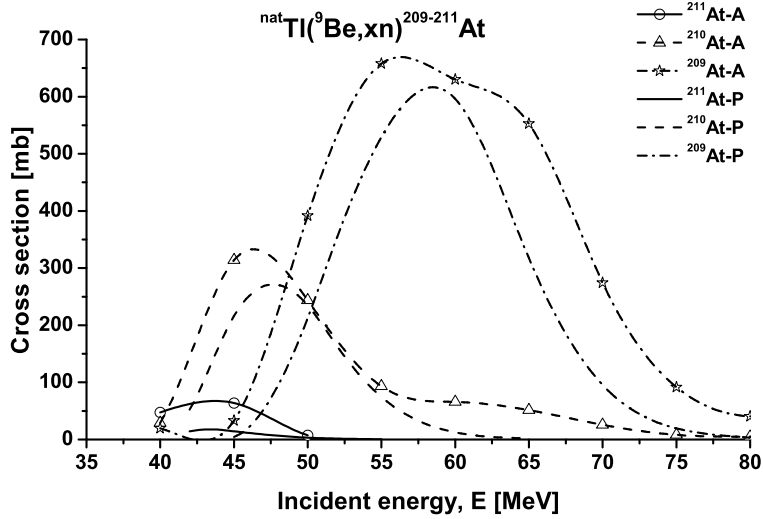}
\caption{Comparison between the formation cross sections of $^{209-211}$At from $^{9}$Be-induced reaction on $^{nat}$Tl target using ALICE91 and PACE-II.} 
\label{fig6}
\end{center}
\end{figure}

Though the favored production route of $^{209-211}$At is from $\alpha$-induced reaction on Bi target, it is also possible to produce $^{209-211}$At using light heavy ion induced reaction. In this report, we have found possible production routes of $^{209-211}$At from $^{6,7}$Li and $^{9}$Be induced reactions on $^{nat}$Pb and $^{nat}$Tl targets.  Figure \ref{fig4}-\ref{fig6} represent the comparative study of the formation cross section of $^{209-211}$At produced through ($^{7}$Li, $xn$), ($^{6}$Li, $xn$) and ($^{9}$Be, $xn$) reactions on $^{nat}$Pb and $^{nat}$Tl targets using the codes ALICE91 and PACE-II. 

The four naturally occurring isotopes of Pb are $^{208}$Pb (52.4\%), $^{207}$Pb (22.1\%), $^{206}$Pb (24.1\%), $^{204}$Pb (1.4\%). Tl has two naturally occuring isotopes, $^{205}$Tl (70.5\%) and $^{203}$Tl (29.5\%). The formation cross sections have been calculated separately from each isotope of Pb using $^{7}$Li and $^{6}$Li projectiles and that from $^{9}$Be induced reactions on Tl isotopes. The total formation cross section is calculated taking the weighted average of all the naturally occulting isotopes. The same procedure has been followed for the results obtained from both the codes ALICE91 and PACE-II.

The total formation cross section obtained from ALICE91 is the sum of PEQ and EQ cross sections. In case of PEQ calculation in the framework of hybrid model for heavy ion induced reaction we have considered the initial exciton number as the total number of the nucleons in the projectile. The EQ cross section is calculated using the Weisskopf-Ewing formalism.  The formation cross sections obtained from PACE-II is only from the evaporation calculation following the Hauser-Feshbach formalism.
 
In figure \ref{fig4} and \ref{fig5}, production cross sections of $^{209-211}$At obtained from ALICE91 are larger than that obtained from PACE-II but the maximum production cross section is obtained at a particular incident energy for both the model calculations.  Production of $^{211}$At is not significant in the $^{6}$Li-induced reaction on $^{nat}$Pb target. However, $^{209-210}$At will be produced in good amount in both $^{7}$Li and $^{6}$Li induced reaction on natural Pb target.
Figure \ref{fig6} shows the production cross section of $^{209-211}$At radionuclides from $^{9}$Be induced reaction on $^{nat}$Tl target. Almost no production of $^{211}$At is seen in this case with sufficiently good production of two other radionuclides. In all these three cases (figure \ref{fig4}-\ref{fig6}), ALICE91 prediction is larger than PACE-II. This may be due to the added contribution from PEQ reactions in ALICE91.

\begin{figure}
\begin{center}
\includegraphics[height=8.cm]{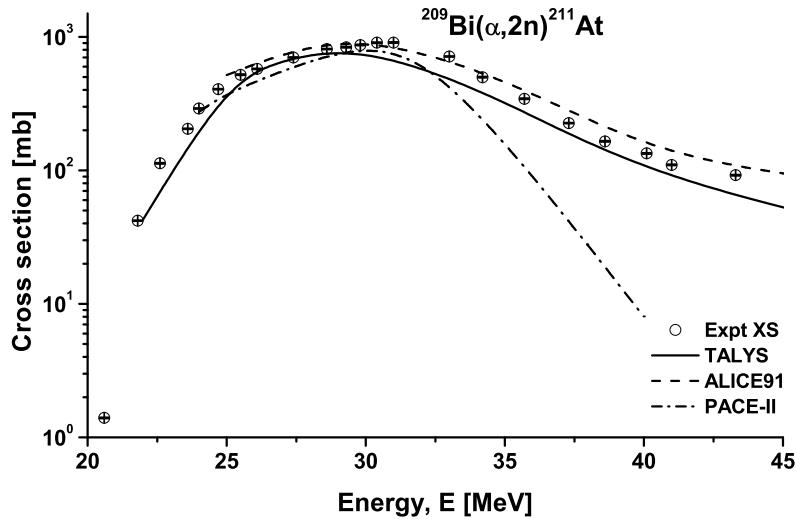}
\caption{Comparison of measured formation cross section \cite{6} of $^{211}$At from $\alpha$-induced reaction on $^{209}$Bi with the TALYS, ALICE91 and PACE-II predictions.} 
\label{fig7}
\end{center}
\end{figure}

\begin{figure}
\begin{center}
\includegraphics[height=8.cm]{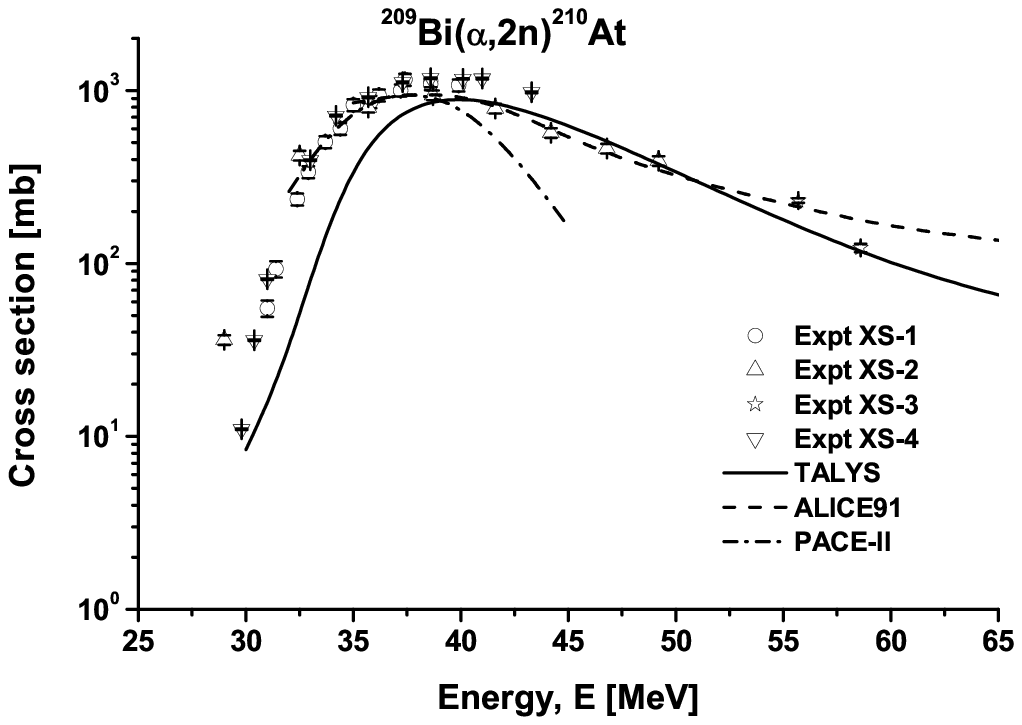}
\caption{Same as figure \ref{fig7} for $^{210}$At \cite{1, 38,39,6}} 
\label{fig8}
\end{center}
\end{figure}

In figure \ref{fig7}-\ref{fig8}, we have compared the measured production cross sections of $^{211}$At \cite{6} and $^{210}$At \cite{1, 38, 39, 6} produced from $\alpha$-induced reactions on $^{209}$Bi target through ($\alpha$, $2n$) and ($\alpha$, $3n$) channels reported by several authors with the theoretical predictions obtained from TALYS, ALICE91 and PACE-II. Figure \ref{fig7} shows that TALYS perfectly matches with the measured data of $^{211}$At \cite{6} in the range shown. ALICE91 also matches with the measured data except in the low energies. PACE-II reproduces the data up to 32 MeV.  It underpredicts the data showing a abrupt fall in the cross section above 32 MeV. This comparison reveals that PEQ reaction mechanism plays an important role in reproducing the measured data. We have seen in figure \ref{fig2} that in the energy range up to 100 MeV, DIR reaction has absolutely no role. As TALYS and ALICE91 account the PEQ reactions along with EQ reaction, they produce the measured data satisfactorily.  Figure \ref{fig8} shows that TALYS also evaluates the measured formation cross section of $^{210}$At. ALICE91 produces the data well but not below 32 MeV. PACE-II shows the same nature as the production of $^{211}$At.

\begin{figure}
\begin{center}
\includegraphics[height=8.cm]{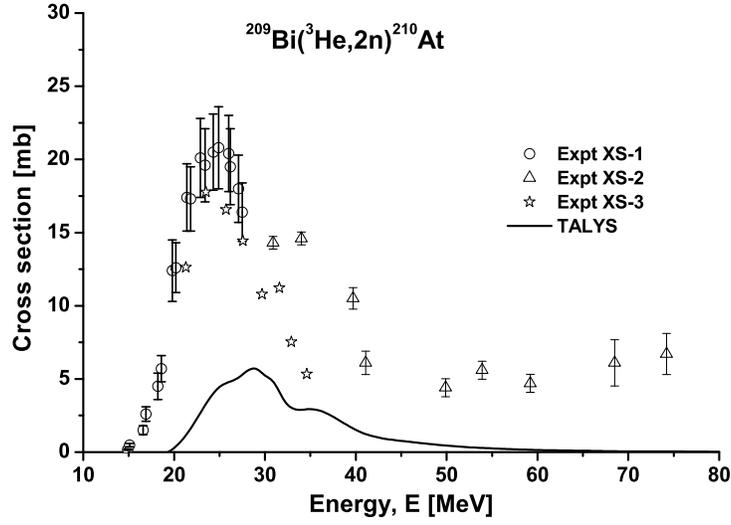}
\caption{Comparison of measured formation cross section of $^{210}$At from $^{3}$He-induced reaction \cite{42, 43, 44} on $^{209}$Bi using TALYS.} 
\label{fig9}
\end{center}
\end{figure}

\begin{figure}
\begin{center}
\includegraphics[height=8.cm]{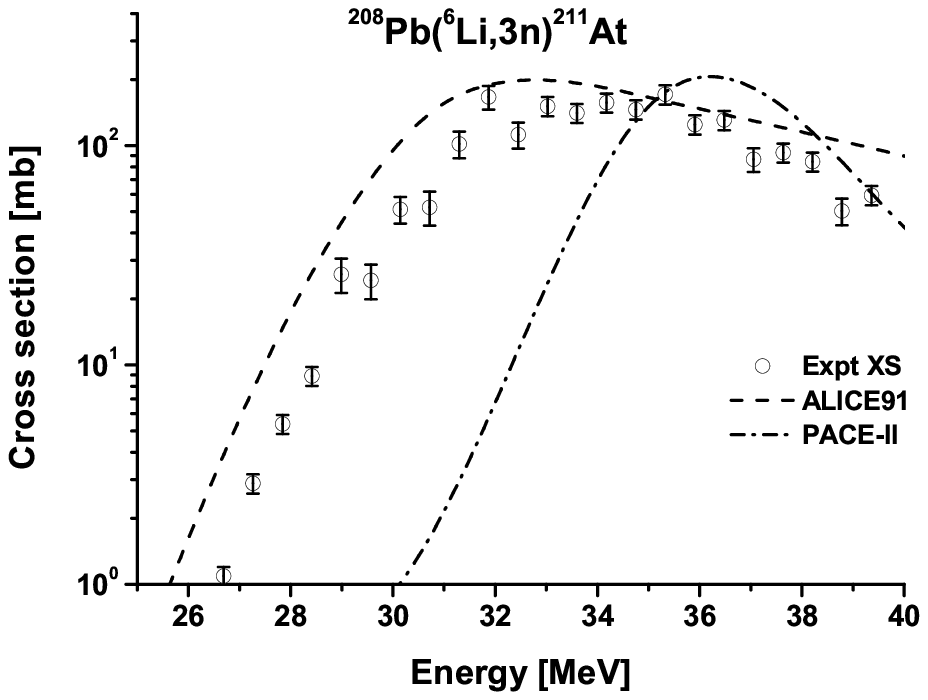}
\caption{ Comparison of measured formation cross section of $^{211}$At from $^{6}$Li-induced reaction \cite{36} on $^{208}$Pb with the ALICE91 and PACE-II calculations.} 
\label{fig10}
\end{center}
\end{figure}

\begin{figure}
\begin{center}
\includegraphics[height=8.cm]{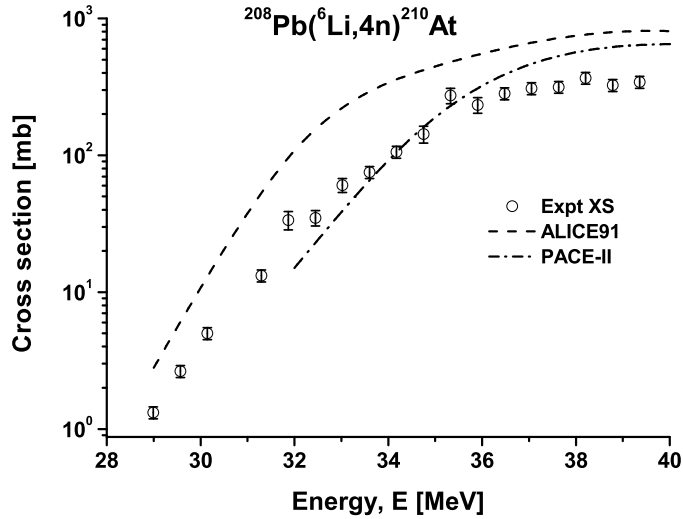}
\caption{Same as figure \ref{fig10} for $^{210}$At \cite{36}. }
\label{fig11}
\end{center}
\end{figure}

Figure \ref{fig9} compares the measured excitation function of $^{210}$At from $^{3}$He-induced reaction \cite{42, 43, 44} on the $^{209}$Bi target with the TALYS predicted value.  The trend of the measured data is well reproduced by TALYS, but its absolute value underpredicts the measured data by the factor of five. ALICE91 absolutely fails to produce the data.
Figure \ref{fig10}-\ref{fig11} present comparison between measured data of $^{211}$At and $^{210}$At produced from $^{208}$Pb($^{6}$Li,3$n$)$^{211}$At \cite{36} and $^{208}$Pb($^{6}$Li,4$n$)$^{210}$At \cite{36} reactions with the theoretical values from ALICE91 and PACE-II. From figure \ref{fig10}, we observe that formation cross section of $^{211}$At is well reproduced by the calculated values of ALICE91 within the uncertainty range. But PACE-II only reproduces the higher energy side of the spectrum. In case of formation cross section of $^{210}$At in figure \ref{fig11}, ALICE91 and PACE-II both reproduce the trend of the measured data. Absolute values of cross section obtained from PACE-II produce the measured data, whereas ALICE91 slightly overpredicts the data.

\begin{figure}
\begin{center}
\includegraphics[height=8.cm]{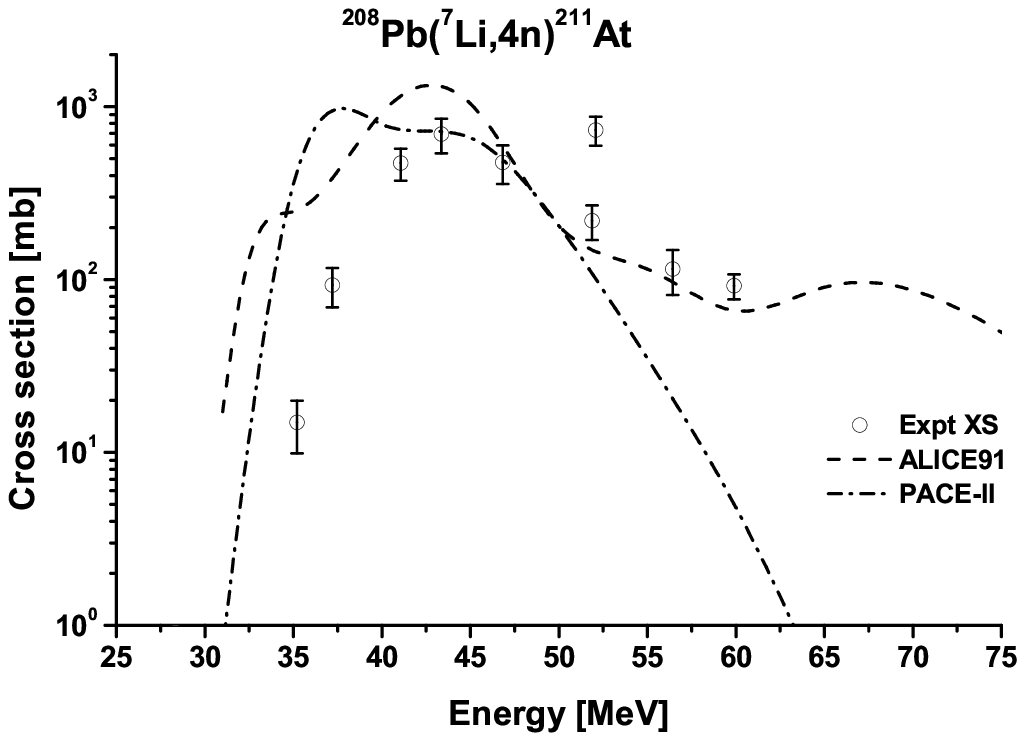}
\caption{Comparison of measured formation cross section of $^{211}$At from $^{7}$Li-induced reaction on $^{208}$Pb \cite{37} with the ALICE91 and PACE-II calculations.} 
\label{fig12}
\end{center}
\end{figure}

\begin{figure}
\begin{center}
\includegraphics[height=8.cm]{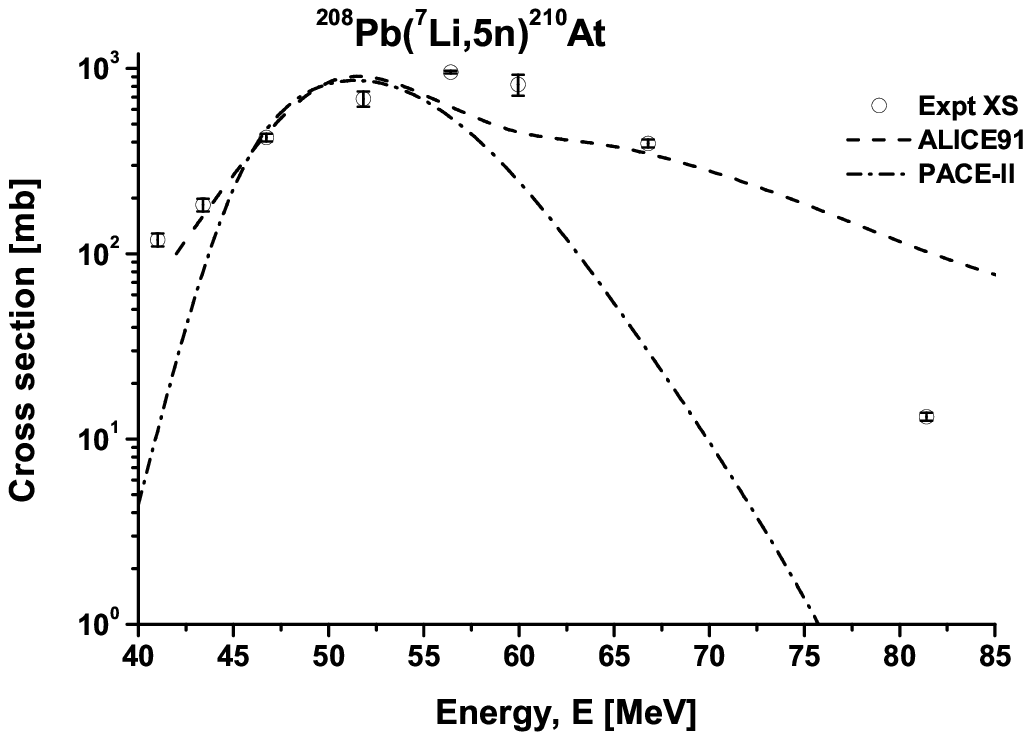}
\caption{Same as figure \ref{fig12} for $^{210}$At \cite{37}.} 
\label{fig13}
\end{center}
\end{figure}

In a similar fashion, we have compared production of $^{211}$At \cite{37} and $^{210}$At \cite{37} from $^{7}$Li-induced reaction on $^{208}$Pb target. The trend of measured cross section of $^{211}$At in figure \ref{fig12} is well reproduced by ALICE91 but absolute value of cross sections over predicts the measured data below 45 MeV. PACE-II neither produces trend or the absolute values of cross section except for a particular energy 45 MeV. In case of figure \ref{fig13}, we see that ALICE91 explains the measured data well but PACE-II under predicts the data mostly.

\section{	Conclusions}
The study of the heavy ion induced production of $^{209-211}$At radionuclides, we can conclude that PEQ reaction mechanism plays an important role in case of light heavy ion induced reactions in the energy range studied and performance of the ALICE91 code is better in most cases. PACE-II mostly underpredicts the measured at high energy side where PEQ reaction contributes. As the code does not consider the PEQ reaction in its calculation, underpredicts the measured data. TALYS and ALICE91 calculated excitation functions of At radionuclides from $\alpha$-induced reaction are comparable and both agree with the experimental data at higher energies, but the cross sections at low energies are well reproduced by TALYS. Heavy ion induced production cross sections of $^{209-211}$At obtained from ALICE91 are mostly in good agreement with the measured data, whereas PACE-II calculations underpredict them. This may be due to the significant contribution from PEQ reactions which is accounted in ALICE91. The study on the contribution of various reaction mechanisms to the total cross section agrees with fact. Direct reaction has no role in the production of At radionuclides from both light as well as heavy ion reactions.

\begin{acknowledgments}
This work has been carried out as a part of  the Saha Institute of Nuclear Physics-Departmet of Atomic Energy, XI five year plan project "Trace Analysis: Detection, Dynamics and Speciation (TADDS)". 
\end{acknowledgments}

\end{document}